\documentclass[submission,copyright,creativecommons]{eptcs}
\providecommand{\event}{ICLP 2022}

\usepackage[utf8]{inputenc}
\usepackage{hyperref}
\hypersetup{final}
\usepackage{hybrid}

\usepackage{todonotes}

\title{An Iterative Fixpoint Semantics for MKNF Hybrid Knowledge Bases with Function Symbols}

\author{Marco Alberti \institute{Dipartimento di Matematica e Informatica, University of Ferrara}
\and Riccardo Zese \institute{Dipartimento di Scienze Chimiche, Farmaceutiche ed Agrarie, University of Ferrara}
\and Fabrizio Riguzzi \institute{Dipartimento di Matematica e Informatica, University of Ferrara}
\and Evelina Lamma \institute{Dipartimento di Ingegneria, University of Ferrara}}

\def\titlerunning{An Iterative Fixpoint Semantics for MKNF Hybrid Knowledge Bases with Function Symbols}
\def\authorrunning{M. Alberti, R. Zese, F. Riguzzi \& E. Lamma}

\begin{document}
\maketitle

\begin{abstract}
  Hybrid Knowledge Bases based on Lifschitz's logic of Minimal Knowledge with Negation as Failure are a successful approach to combine the expressivity of Description Logics and Logic Programming in a single language.
  Their syntax, defined by Motik and Rosati, disallows function symbols.
  In order to define a well-founded semantics for MKNF HKBs, Knorr et al. define a partition of the modal atoms occurring in it, called the alternating fixpoint partition.
  In this paper, we propose an iterated fixpoint semantics for HKBs with function symbols.
  We prove that our semantics extends Knorr et al.'s, in that, for a function-free HKBs, it coincides with its alternating fixpoint partition.
  The proposed semantics lends itself well to a probabilistic extension with a distribution semantic approach, which is the subject of future work.
\end{abstract}

% \begin{keywords}
%   MKNF, Hybrid Knowledge Bases, Function Symbols, Fixpoint Semantics
% \end{keywords}

\section{Introduction}
\label{sec:intro}

When modelling complex domains it is of foremost importance to choose the logic that better fits with what must be represented. Therefore, many languages have been defined, based on First Order Logic such as Logic Programming (LP) or Description Logic (DL). These languages share many similarities but, on the other hand, they differ in the domain closure assumption they make: closed-world assumption for LP and open-world assumption for DLs.

Since many domains, such as legal reasoning \cite{krr-clima-xii}, require different closure assumptions to  coexist in the same model, combinations of LP and DL have been proposed by several authors. One of the most effective approaches is called Minimal Knowledge with Negation as Failure (MKNF)~\cite{DBLP:conf/ijcai/Lifschitz91}.
MKNF was then applied to define hybrid knowledge bases (HKBs)~\cite{MotikRosatiMKNF_J_ACM}, which are defined as the combination of a logic program and a DL KB.

In the original HKB language, function symbols are not allowed.
However,  this is a feature that is useful in many domains.
Consider, for example, the behaviour of a virus, which can mutate and spillover may happen due to each mutation. To trace the evolution of a virus, it is necessary to identify the sequence of spillover events starting from the initial version of the virus. We can represent the spillover count by Peano numbers, by means of a function symbol $s/1$ modelling that, e.g., $s(Y)$ represents the spillover event that follows the spillover identified by $Y$, which may have happened after another spillover, and so forth.

In this paper, we propose to extend the HKB syntax with function symbols,
and we present an iterated fixpoint semantics for HKBs with Function Symbols (\hkbfs).
We prove that our semantics coincides with that of \cite{KnorrMKNF_AI} and \cite{LiuYou_AI2017_HKB_WFS} in the case of HKBs not including function symbols,
and therefore can be considered an extension of that semantics to the case with function symbols.

The proposed semantics will also serve as the basis for a further (probabilistic) extension of the language, based on a distribution semantics approach, which is the subject of an ongoing effort.

We provide the necessary background notions in Section \ref{sec:background}.
We define the syntax and semantics of {\hkbfs}s in Section \ref{sec:hkbs-with-function-symbols}.
In Section \ref{sec:properties}, we prove that our semantics extends Knorr et al's.
We  conclude the paper in Section \ref{sec:conclusions-future-work}.

\section{Background}\label{sec:background}

In this section, we provide the necessary background notions on the syntax and semantics of the language of MKNF Hybrid Knowledge Bases, which we extend with function symbols in Section \ref{sec:hkbs-with-function-symbols}. We start with Description Logics, which are a part  of the language of HKBs.
% Preliminary definitions about FOL and Logic Programming   are in \ref{app:background},
% to make the paper self-contained.

\subsection{Description Logics}
\label{sec:bg-dl}

Description Logics (DLs) are decidable fragments of First Order Logic
used to model ontologies \cite{borgida1996relative}.
Usually their syntax is based on concepts and roles,
corresponding  to unary  and binary predicates, respectively.
In the following we briefly recall the DL $\mathcal{ALC}$;
see~\cite{Baader:2003:DLH:885746coll} for a complete introduction to DLs.
%Throughout the paper, we assume that the DLs at hand are decidable,
%i.e., there exists an algorithm able to answer any query in a finite number of steps.

$\mathcal{ALC}$'s alphabet is composed of
a set $\mathbf{C}$ of \emph{atomic concepts},
a set $\mathbf{R}$ of \emph{atomic roles}
and a set $\mathbf{I}$ of individuals.
A \emph{concept} $C$ is defined by:
\begin{align}
  C::= & C_1|\bot|\top|(C\sqcap C)|(C\sqcup C )|\neg C|\exists R.C|\forall R.C\notag
\end{align}
where $C_1\in\mathbf{C}$ and $R\in\mathbf{R}$.

A \emph{TBox} $\cT$ is a finite set of
\textit{concept inclusion	axioms} $C\sqsubseteq D$,
where $C$ and $D$ are concepts.
An \emph{ABox} $\cA$ is a finite set of
\textit{concept membership 	axioms} $a : C$
and \textit{role membership axioms} $(a, b) :	R$,
where $C$ is a concept, $R \in \mathbf{R}$ and $a, b \in \mathbf{I}$.
An $\mathcal{ALC}$ knowledge base $\cK = (\cT , \cA)$
consists of a TBox $\cT$ and an ABox $\cA$.

DL axioms can be mapped to FOL formulas by the transformation $\pi$
shown in Table~\ref{table:translation} for the $\mathcal{ALC}$ DL
\cite{DBLP:conf/dlog/Sattler03}. $\pi$
is applied to concepts as follows:
\[
  \begin{array}{rcl}
    \pi_x(A)           & = & A(x)                                \\
    \pi_x(\neg C)      & = & \neg \pi_x (C)                      \\
    \pi_x(C\sqcap D)   & = & \pi_x(C)\wedge\pi_x(D)              \\
    \pi_x(C\sqcup D)   & = & \pi_x(C)\wedge\pi_x(D)              \\
    \pi_x(\exists R.C) & = & \exists y.R(x,y)\wedge\pi_y(C)      \\
    \pi_x(\forall R.C) & = & \forall y.R(x,y)\rightarrow\pi_y(C) \\
  \end{array}
\]

\begin{table}
  \begin{center}
    \begin{tabular}{c|c}
      Axiom                   & Translation                              \\
      \hline $C\sqsubseteq D$ & $\forall x.\pi_x(C)\rightarrow \pi_x(D)$ \\
      $a : C$                 & $\pi_a(C)$                               \\$(a,b):R$&$R(a,b)$\\
      $a =b$                  & $a=b$                                    \\
      $a\neq b$               & $a\neq b$
    \end{tabular}
  \end{center}
  \caption{Translation of $\mathcal{ALC}$  axioms into FOL.  }
  \label{table:translation}
\end{table}

\subsection{MKNF-based Hybrid Knowledge Bases}\label{sec:hkb}

The logic of Minimal Knowledge with Negation as Failure (MKNF)
was introduced in~\cite{DBLP:conf/ijcai/Lifschitz91} to support epistemic queries on logic programs.
MKNF was inspired by several works
\cite{DBLP:journals/ai/Levesque84,Reiter1990}
on epistemic query answering on non-monotonic databases,
which is essential when databases contain incomplete information.

The syntax of MKNF
is the syntax of FOL
augmented with the modal operators $\mk$ and $\mnot$.

MKNF-based Hybrid Knowledge Bases ~\cite{MotikRosatiMKNF_J_ACM} are combinations of DL axioms and LP rules
that can be mapped to a MKNF formula, as follows. As shown in~\cite{MotikRosatiMKNF_J_ACM},
MKNF-based HKBs exhibits desirable properties
(faithfulness, i.e., preservation of the semantics of both formalisms when the other
is absent;
tightness, i.e., no layering of LP and DL;
flexibility, i.e., the possibility to view each predicate under both open and closed world assumption;
decidability),
which each of the other existing approaches
to LP and DL integration lacks at least partly.

\begin{definition}\label{def:hkb}
  A Hybrid Knowledge Base (HKB) is a pair $\kb = \kbpair$
  where $\ont$ is a set of axioms in a %decidable 
  description logic  (Section~\ref{sec:bg-dl})
  and $\prog$ is a finite set of normal function-free logic programming rules.

\end{definition}
In the rest of the paper, with a slightly abuse of notation, we will say that a HKB $\kb_1=\langle \ont_1, \prog{}_1 \rangle$ is a subset of a HKB $\kb_2=\langle \ont_2, \prog{}_2 \rangle$, i.e., $\kb_1\subseteq\kb_2$ iff $\ont_1\subseteq\ont_2$ and $\prog{}_1\subseteq\prog{}_2$.
Given a HKB $\kb=\kbpair$,
an atom in \prog is a \emph{DL-atom}
if its predicate occurs in $\ont$, a non-DL-atom otherwise.

\begin{definition}[DL-safety]
  \label{def:hkb-dl-safety}
  A rule is \emph{DL-safe}
  if each of its variables occurs in at least one positive non-DL-atom in the body;
  a HKB is \emph{DL-safe}
  if all its rules are DL-safe.
\end{definition}

In this paper, we assume that all HKBs are DL-safe.

An HKB $\kb = \kbpair$ can be mapped to an MKNF formula
by extending the standard transformation \Transform for DL axioms (Table \ref{table:translation})
to support LP rules:
\begin{itemize}
  \item if $r$ is a rule of the form
        $h \lpif a_1, \ldots, a_n,\lpnot b_1, \ldots, \lpnot b_m$ where all $a_i$ and $b_j$ are  atoms and
        $\mathbf{X}$ is the tuple of all variables in $r$, then
        $\transform{r} = \forall \mathbf{X} (\mk a_1 \wedge \ldots \wedge
          \mk a_n \wedge \mnot b_1 \wedge \ldots \wedge \mnot b_m
          \lthen \mk h)$
  \item $\transform{\prog} = \bigwedge_{r \in \prog} \transform{r}$
  \item $\transform{\kbpair} = \mk \transform{\ont} \wedge \transform{\prog}$
\end{itemize}

\noindent This transformation is a way to give a semantics to a HKB:
MKNF formulas have been given
two-valued~\cite{MotikRosatiMKNF_J_ACM}
and three-valued \cite{KnorrMKNF_AI} semantics,
so the (two or three-valued) semantics of the resulting MKNF formula
can be taken as the semantics of the original HKB.
We refer the reader to those articles for an in-depth discussion of the semantics and their respective merits.

In the following, we recall the three-valued MKNF semantics,
which is more relevant to our work.
For simplicity, we omit the signature \signature from the definitions.

\paragraph{Three-valued MKNF semantics~\cite{KnorrMKNF_AI}}

The truth of an MKNF formula $\psi$ is defined relatively to
a \emph{three-valued MKNF structure} $(I, \mathcal{M},\mathcal{N})$,
which consists of a first-order interpretation $I$ over a universe \universe\
and two pairs $\mathcal{M} = (M,M_1)$ and $\mathcal{N} = (N,N_1)$ of sets of
first-order interpretations over \universe where $M_1 \subseteq M$ and
$N_1 \subseteq N$.
$\mk \psi$ is true (resp. false) with respect to $(M,M_1)$ if and only if $\psi$ is true in all elements of $M$ (resp. not true in all elements of $M_1$).
$N$ and $N_1$ serve the same purpose for defining the truth value of $\mnot \psi$.

Satisfaction of a closed formula by a three-valued
MKNF structure is defined as follows
(where $p$ is a predicate,
$\psi$ is a formula,
the values \true, \undefine and \false follow the order $\false<\undefine<\true$,
and $\epsilon^I$ represents the individual or relation in the domain of discourse assigned to $\epsilon$ by the interpretation $I$):
% \hspace*{0.5cm}

\begin{small}
  $$
    \begin{array}{ll}
      (I,\mathcal{M},\mathcal{N})(p(t_1,\ldots,t_n))    & \true\ \sse\
      (t_1^I,\ldots,t_n^I)
      \in p^I                                                                                                                                        \\
                                                        & \false\ \sse\
      (t_1^I,\ldots,t_n^I)
      \not\in p^I                                                                                                                                    \\

      (I,\mathcal{M},\mathcal{N})(\neg \psi)            & \true\ \sse\ (I,\mathcal{M},\mathcal{N}) (\psi)  = \false,                                 \\ & \undefine\ \sse\ (I,\mathcal{M},\mathcal{N}) (\psi)  = \undefine,\\ & \false\ \sse\ (I,\mathcal{M},\mathcal{N}) (\psi)  = \true\\
      (I,\mathcal{M},\mathcal{N})(\psi_1 \wedge \psi_2) & min\{(I,\mathcal{M},\mathcal{N})(\psi_1), (I,\mathcal{M},\mathcal{N})(\psi_2)\}            \\
      (I,\mathcal{M},\mathcal{N})(\psi_1 \lthen \psi_2) & \true\ \sse\ (I,\mathcal{M},\mathcal{N})(\psi_1) \leq (I,\mathcal{M},\mathcal{N})(\psi_2), \\
                                                        & \false\ \textnormal{ otherwise}                                                            \\
      (I,\mathcal{M},\mathcal{N})(\exists x: \psi)      & max\{(I,\mathcal{M},\mathcal{N})(\psi[\alpha/x])| \alpha \in \universe\}                   \\
      (I,\mathcal{M},\mathcal{N})(\mk \psi)             & \true\ \sse\ (J,(M,M_1),\mathcal{N}) (\psi)=\true\ \textnormal{ for all } J \in M,         \\
                                                        & \false\ \sse\ (J,(M,M_1),\mathcal{N}) (\psi)=\false\  \textnormal{ for some } J \in M_1,   \\
                                                        & \undefine \textnormal{ otherwise}                                                          \\
      (I,\mathcal{M},\mathcal{N})(\mnot \psi)           & \true\ \sse\ (J,\mathcal{M},(N,N_1)) (\psi)=\false \textnormal{ for some } J \in N_1,      \\
                                                        & \false\ \sse\ (J,\mathcal{M},(N,N_1)) (\psi)=\true\ \textnormal{ for all } J \in N,        \\
                                                        & \undefine \textnormal{ otherwise}
    \end{array}$$
\end{small}
An \emph{MKNF interpretation} over a universe \universe is a non-empty
set of first order interpretations over \universe.  An \emph{MKNF
  interpretation pair} $(M, N)$ over a universe \universe consists of
two MKNF interpretations $M$, $N$ over \universe, with
$\emptyset \subset N \subseteq M$.
An MKNF interpretation pair  $(M, N)$
\emph{satisfies} a closed MKNF formula $\psi$ iff,
for each $I \in M$,
$(I,(M,N),(M,N))(\psi) = \true$.
If $M = N$, then the MKNF interpretation pair $(M, N)$ is called \emph{total}.  If there exists an MKNF interpretation pair satisfying $\psi$,
then $\psi$ is \emph{consistent}.
An MKNF interpretation pair $(M, N)$ over a universe
\universe is a \emph{three-valued MKNF model} for a given closed MKNF
formula $\psi$ if
\begin{itemize}
  \item $(M, N)$ satisfies $\psi$ and
  \item for each MKNF interpretation pair $(M',N')$ over $\universe$ with
        $M \subseteq M'$ and $N \subseteq N'$, where at least one of the
        inclusions is proper and $M' = N'$ if $M = N$, there is $I' \in M'$
        such that $(I',(M',N'),(M,N))(\psi) = \false$.
        In other words,
        $M$ and $N$ cannot be extended while satisfying $\psi$;
        the semantics implements minimal knowledge
        by requiring as many possible worlds as possible.
\end{itemize}

\subsection{Well Founded HKB Semantics}
\label{sec:bg-hkb-wfs}

% The semantics of an HKB is defined by mapping it into a formula in the logic of Minimal Knowledge with Negation as Failure \cite{DBLP:conf/ijcai/Lifschitz91}; the HKB's MKNF models are the models of the MKNF formula. The details are in Appendix \ref{app:MKNF-semantics}.

In \cite{KnorrMKNF_AI}, the well-founded model of an MKNF formula is defined as the three-valued MKNF model that, intuitively, leaves as much as possible undefined.
% In particular, the authors define a ``more knowledge
% derivable" relation between MKNF interpretation pairs:
% $(M_1,N_1) \knowge (M_2,N_2)$ iff $M_1 \subseteq M_2$ and
% $N_2 \subseteq N_1$.  An HKB's three-valued MKNF model $(M,N)$ that is
% minimal w.r.t. \knowge\ (i.e., if $(M_1,N_1)$ is also a three-valued
% model, then $(M_1,N_1) \knowge (M,N)$) is defined to be a \emph{well-founded
% model}.
Not all HKBs have a well-founded model;
\emph{MKNF-coherent} HKBs \cite{LiuYou_AI2017_HKB_WFS}
have a unique well-founded model
that is characterized by a partition of the atoms that occur in rules,
called the alternating fixpoint partition and defined by \cite{KnorrMKNF_AI}.

The NoHR query answering system \cite{DBLP:journals/ki/KasalicaKLL20} is based on the well-founded semantics for HKBs.

We recall these definitions below.

An MKNF formula $\psi$ is ground if $\psi$ does not contain variables.
Given a hybrid MKNF knowledge base $\kb = \kbpair$, the ground instantiation of $\kb$ is the KB $\kb_g = \langle \ont, \prog{}_g \rangle$ where
$\prog{}_g$ is obtained from $\prog$ by replacing each rule $r$ of $\prog$ with a set of rules substituting each variable in $r$ with constants from 	$\kb$ in all possible ways.
Let $\kb = \kbpair$ be a ground HKB.
Note that, if an HKB is DL-safe, it has the same two-valued \cite{MotikRosatiMKNF_J_ACM} and three-valued \cite{KnorrMKNF_AI} MKNF models of its grounding over the constants that occur in it,
so it can be assumed, without loss of generality, that the HKB is ground.
The set of \emph{known atoms} of
\kb, \ka{\kb}, is the set of all (ground) atoms occurring in $\prog$ \cite{MotikRosatiMKNF_J_ACM}.

\begin{definition}
  \label{def:ka-partition}
  A \emph{partition} of \ka{\kb} is a pair $(P,N)$ such that
  $P \subseteq N \subseteq \ka{\kb}$; $(P,N)$ is \emph{exact} if $P=N$.
\end{definition}
Given $S \subseteq \ka{\kb}$, the \emph{objective knowledge of \ont
  with respect to $S$} is the set of first order formulas
\begin{equation}
  \label{eq:objective-knowledge}
  \obknowl{\kb}{S} = \{ \transform{\ont} \} \cup S
\end{equation}
where $\pi$ is the standard transformation \Transform for DL axioms (Table \ref{table:translation}).

The operators $\RKop{\kb}$, $\DKop{\kb}$ and $\TKop{\kb}$
derive atoms that are
consequences of a positive HKB $\kb$
(i.e., one where no negative literals occur in rules)
and a set $S$ of atoms.
$\RK{\kb}{S}$ is the set of immediate consequences due to rules,
i.e., the heads of rules in $\prog$ whose bodies are composed of atoms that are a subset of $S$;
$\DK{\kb}{S}$ is the set of immediate consequences due to axioms,
i.e., the atoms from \ka{\kb} entailed by \obknowl{\kb}{S};
and $\TK{\kb}{S} = \RK{\kb}{S} \cup \DK{\kb}{S}$.
Given an HKB $\kb$ and a
set of atoms $S \subseteq \ka{\kb}$, the following transformations,
which yield positive knowledge bases, are defined: the \emph{MKNF
  transformation} $\mknftrans{\kb}{S}$ is
$\pair{\ont}{\mknftrans{\prog}{S}}$ where $\mknftrans\prog{S}$
is the set of rules $h \lpif a_1,\ldots,a_m$ such that there exists in
\prog a rule
$h \lpif a_1,\ldots,a_m, \lpnot b_1, \ldots , \lpnot b_n$ with
$\{ b_1,\ldots, b_n\} \cap S = \emptyset$, and the \emph{MKNF-coherent
  transformation} $\mknfcohtrans\kb{S}$ is
$\pair\ont{\mknfcohtrans{\prog}{S}}$ where
$\mknfcohtrans{\prog}{S}$ is the set of rules $h \lpif a_1,\ldots,a_m$
such that there exists in $\prog$ a rule
$h \lpif a_1,\ldots,a_m, \lpnot b_1, \ldots , \lpnot b_n$ with
$\{b_1,\ldots, b_m\} \cap S = \emptyset$ and
$\obknowl{\kb}{S} \not\models \neg h$.

Since, as shown in \cite{KnorrMKNF_AI}, $T_{\kb}$ is monotonic if
$\kb$ is a ground positive HKB, the following transformations of sets
of atoms are well defined:
$\trans{\kb}{S} = \lfp{\TKop{\mknftrans{\kb}{S}}}$ and
$\transp{\kb}{S} = \lfp{\TKop{\mknfcohtrans{\kb}{S}}}$.
Using these transformations, it is possible to define a partition of $\kb$'s known atoms as follows.
\begin{definition}\label{def:alternating-fixpoint-partition}
  For an HKB $\kb$,
  the sequences of sets of atoms $\posatoms{}$ and $\negatoms{}$ are defined as follows:
  $\posatoms{0} = \emptyset$,
  $\negatoms{0} = \ka{\kb}$,
  $\posatoms{n+1} = \trans{\kb}{\negatoms{n}}$ and
  $\negatoms{n+1} = \transp{\kb}{\posatoms{n}}$,
  $\posatomsf = \bigcup \posatoms{i}$,
  $\negatomsf = \bigcap \negatoms{i}$.

  The pair $(\posatomsf,\negatomsf)$ is called $\kb$'s \emph{alternating fixpoint partition}.
\end{definition}

\cite{LiuYou_AI2017_HKB_WFS} identify the class of \emph{MKNF-coherent} HKBs, i.e., those whose alternating fixpoint partition defines a well-founded model,
as well as some sufficient conditions for a HKB to be MKNF-coherent.

We assume that the HKBs that we consider are MKNF-coherent.

\begin{definition}[MKNF-coherent HKB (Def. 10 of
    \cite{LiuYou_AI2017_HKB_WFS})]
  \label{def:mknf-coh}
  An HKB $\kb$ is \emph{MKNF-coherent} if $(I_P, I_N)$, where
  $I_P = \{ I \suchthat I \models \obknowl{\kb}{\posatomsf} \}$ and
  $I_N = \{ I \suchthat I \models \obknowl{\kb}{\negatomsf} \}$, is a
  three-valued MKNF model of \kb.
\end{definition}

For MKNF-coherent HKBs, the model determined by the alternating
fixpoint partition as in Definition \ref{def:mknf-coh} is the unique
well-founded model.

\begin{proposition}[Proposition 2 of
    \cite{LiuYou_AI2017_HKB_WFS}]\label{thm:unique-wfm}
  If $\kb$ is an MKNF-coherent HKB, then it has the unique well-founded
  model
  $(\{ I \suchthat I \models \obknowl{\kb}{\posatomsf} \},\{ I
    \suchthat I \models \obknowl{\kb}{\negatomsf} \})$
\end{proposition}

We report some sufficient conditions for MKNF-coherence from \cite{LiuYou_AI2017_HKB_WFS} in Appendix \ref{app:MKNF-coherent-hkbs}.

Intuitively, the alternating fixpoint partition marks each known atom in $\ka\kb$ and induces the well founded model.

% For an MKNF-coherent HKB $\kb$ with alternating fixpoint partition $(\posatomsf,\negatomsf)$ and $a \in \ka{\kb}$, we write $\WFM(\kb) \models a$ if $a \in \posatomsf$ and $\WFM(\kb) \models \neg a$ if $a \in \ka{\kb} \setminus \negatomsf$.

\section{HKBs with function symbols}
\label{sec:hkbs-with-function-symbols}

In this section, we extend the language of HKBs
(Section~\ref{sec:hkb})
to allow  function symbols.
We define the syntax in Section \ref{sec:hkbfs-language} and the semantics in Section \ref{sec:hkbfs-semantics}.
We also provide a running example (Example \ref{ex:mutanp}) of the proposed syntax and semantics,
which takes advantage of function symbols to model natural numbers.

\subsection{Language}
\label{sec:hkbfs-language}

The syntax extension amounts to lifting the function-free limitation of the original HKB syntax.

\begin{definition}[Hybrid Knowledge Base with Function Symbols]\label{def:hkbfs}
  A Hybrid Knowledge Base with Function Symbols (\hkbfs)
  is a Hybrid Knowledge Base (Section~\ref{sec:hkb})
  whose rules can contain function applications.
\end{definition}

The definition of DL-safety (Def. \ref{def:hkb-dl-safety}) also applies to {\hkbfs}s.
In this paper, we assume that all {\hkbfs}s are DL-safe.

% \todo[inline]{Esempio di Riccardo}

\begin{example}[Spillover]
	\label{ex:mutanp}
	Let $\kb = \kbpair$, where
	\begin{align}
		\prog =\ & \mi{safe}(X) \lpif \lpnot{\mi{spillover\_count}(X,\mi{s}(\mi{s}(Y)))}. \nonumber  \\
		& \mi{spillover\_count}(X, \mi{s}(Y)) \lpif  \mi{virus}(X), \mi{mutated}(X), 
		%  \nonumber  \\
		% &\>\>\>\>\>\>\>\>\>\>\>\>\>\>\>\>\>\>\>\>\>\>\>\>\>\>\>\>\>\>\>\>\>\>\>\>\>\>\>\>\>\>\>\>\>\>\>\>\>\>\>\>\>\>\>\>\>\>\>\>\>\>\>\>\>
		\mi{spillover\_count}(X,Y). \nonumber  \\
		& \mi{spillover\_count}(X,0) \lpif \mi{virus}(X). \nonumber  \\
		& \mi{virus}(t).  \nonumber  \\
		\ont =\  & \exists \mi{mutation}.\top \sqsubseteq \mi{mutated} \nonumber \\
		& \mi{t}: \exists \mi{mutation}.\top  \nonumber
	\end{align}
	
	This HKB models that $\mi{t}$ is a virus and there is at least a mutation of the virus $\mi{t}$. 
	If there exists at least one mutation for virus $\mi{t}$, it is mutated, and so, a spillover may have happened. Finally, we can model the series of spillover events by means of predicate $\mi{spillover\_count}$. Function $\mi{s}(Y)$ represents the successor of $Y$. Finally, a virus is safe if the spillover count is less than two.

%==========================================	
	
	% \noindent
	% The MKNF transformation of this HKB is: \RZ{Serve?}
	% \begin{equation*}
	% 	\begin{array}{l}
	% 		\mk \transform{\ont} \wedge \transform{\prog} =															\\
	% 		\mk (\exists Y:(\mi{mutation}(\mi{t},Y)) \wedge															\\
	% 		\ \ \ \ \forall X:(\mi{mutated}(X)\lpif\exists Y:(\mi{mutation}(X,Y)))) \wedge							\\
	% 		\forall X,Y:(\mk\mi{virus}(X)\wedge\mi{mutated}(X)\supset\mk\mi{spillover\_count}(X, \mi{s}(Y)))\wedge		\\
	% 		\forall X:(\mk\mi{virus}(X)\supset\mk\mi{spillover\_count}(X, 0))\wedge									\\
	% 		\mk\mi{virus}(\mi{t})
	% 	\end{array}
	% \end{equation*} 

%==========================================	

\end{example}

\subsection{Iterated fixpoint \hkbfs semantics }\label{sec:hkbfs-semantics}

In this section,
we define the semantics of an \hkbfs 
as a partition of its known atoms.

We proceed in a bottom-up way, similarly to~\cite{Przy89d}.
In particular,
we define two inner operators (Def.~\ref{def:hkbfs-inner-operators})
that, assuming sets of true and false atoms (a 3-valued interpretation for the \hkbfs, Def.~\ref{def:hkbfs-3-valued-interpretation})
possibly derive new true and false atoms, respectively.
These operators are monotonic in their argument (Proposition~\ref{thm:opTrue-opFalse-monotonic}), so they have a least and a greatest fixpoint,
which are used to define the outer operator (Def.~\ref{def:hkbfs-ifp}) which updates the 3-value interpretation.
The outer operator is itself monotonic (Proposition~\ref{thm:ifp-monotonic}),
so it has a least fixpoint,
which we define (Definition~\ref{def:iterated-fixpoint-semantics})  as the semantics of the \hkbfs.

\newcommand{\Tr}{\mi{Tr}}
\newcommand{\Fa}{\mi{Fa}}
\newcommand{\It}{I_{\mathrm{T}}}
\newcommand{\If}{I_{\mathrm{F}}}

\begin{definition}\label{def:hkbfs-3-valued-interpretation}
  A 3-valued interpretation for an \hkbfs $\kb$
  is a pair $\pair\It\If$
  where $\It$ and $\If$ are disjoint sets of $\kb$'s known atoms,
  i.e., $\It \subseteq \ka\kb$, $\If \subseteq \ka\kb$, $\It \inters \If = \emptyset$

  Given a 3-valued interpretation $\pair\It\If$,
  an atom $a$ is true in it if $a \in \It$,
  false in it if $a \in \If$,
  undefined in it otherwise.

  We also define $\pair{\It}{\If} \le \pair{\It'}{\If'}$ iff $\It \subseteq \It'$ and $\If \subseteq \If'$.
\end{definition}

We denote by $\ThreeVInt\kb$  the set of 3-valued interpretations for an \hkbfs $\kb$.

\begin{definition}\label{def:hkbfs-inner-operators}
  Given a ground \hkbfs $\kb=\kbpair$,
  and a 3-valued interpretation
  $\cI = \pair\It\If$ for $\kb$,
  we define the operators
  $\optrue{\kb}{\cI}: \powerset{\ka\kb} \rightarrow  \powerset{\ka\kb}$
  and
  $\opfalse{\kb}{\cI}: \powerset{\ka\kb} \rightarrow \powerset{\ka\kb}$
  as
  \begin{itemize}
    \item $\optrue{\kb}{\cI}(\Tr) =
            \{a\in\ka\kb \suchthat$
          there is a clause
          $a \leftarrow a_1,...,a_n,\lpnot b_1,\ldots,\lpnot b_r$ in the grounding of $\prog$
          such that for every $i$ ($1 \leq i \leq n$)
          $a_i$ is true in $\cI$ or $a_i\in \Tr$,
          and for every $j$ ($1 \leq j \leq r$) $b_j$ is false in $\cI \}
            \cup
            \{a\in \ka{\kb}|
            \obknowl{\kb}{I_T\cup Tr}\models a\}$;

    \item $\opfalse\kb{\cI}(\Fa) = $
          $\{a \in \ka\kb \suchthat$
          $\obknowl{\kb}{I_T} \models \neg
            a$,
          or, for every clause
          $a \leftarrow a_1,...,a_n,\lpnot b_1,\ldots,\lpnot b_r$ in the grounding of $\prog$,
          there is some $i$ $(1 \leq i \leq n)$ such that $a_i$ is false in $\cI$ or $a_i\in \mathit{Fa}$,
          or there is some $j$ $(1 \leq j \leq r)$ such that $b_j$ is true in $\cI \}
            \cap
            \{a \in \ka{\kb}|
            \obknowl{\kb}{\ka{\kb}\setminus(\If\cup\mathit{Fa})}\not\models a\} $
  \end{itemize}
\end{definition}

In words,
$\optrue{\kb}{\cI}(\Tr)$ represents the true atoms that can be derived from $\kb$ knowing $\cI$ and true atoms $\Tr$,
while $\opfalse{\kb}{\cI}(\Fa)$  represents the false atoms that can be derived from $\kb$ by knowing $\cI$ and false atoms $\Fa$.

\begin{proposition}
  \label{thm:opTrue-opFalse-monotonic}
  Given an \hkbfs $\kb$ and a 3-valued interpretation $\cI$ for $\kb$,
  $\optrue\kb{\cI}$ and $\opfalse\kb{\cI}$ are both monotonic in their argument.

  \begin{proof}
    Monotonicity of $\optrue{\kb}{\cI}$ means that
    if $\Tr\subseteq\Tr'$,
    then $\optrue{\kb}{\cI}(\Tr)\subseteq \optrue{\kb}{\cI}(\Tr')$.
    Analogously, monotonicity of $\opfalse{\kb}{\cI}$ means that
    if $\Fa\subseteq\Fa'$,
    then $\opfalse{\kb}{\cI}(\Fa)\subseteq \opfalse{\kb}{\cI}(\Fa')$.

    Regarding $\optrue{\kb}{\cI}$,
    if $a\in\optrue{\kb}{\cI}(\Tr)$,
    Definition \ref{def:hkbfs-inner-operators} ensures that
    either there is a clause $a \leftarrow a_1,...,a_n,\lpnot b_1,\ldots,\lpnot b_r$ in $\prog$'s grounding
    such that for each $1 \leq i \leq n$
    $a_i$ is true in $\cI$ or $a_i \in \Tr$
    and for each $1 \le j \le r$
    $b_{j}$ is false in $\cI$,
    or $\obknowl{\kb}{\It \cup \Tr}\models a$, i.e., ${\transform{\ont}} \cup  \It \cup \Tr \models a$.
    Since $\Tr\subseteq\Tr'$, if $a_i \in \Tr$, then also $a_i\in \Tr'$,
    and if ${\transform{\ont}} \cup \It \cup \Tr \models a$, then also ${\transform{\ont}} \cup \It \cup \Tr' \models a$ by the monotonicity of first order logic.
    So $a\in\optrue{\kb}{\cI}(\Tr')$.

    Regarding $\opfalse{\kb}{\cI}$,
    if $a\in\opfalse{\kb}{\cI}(\Fa)$,
    then either
    \begin{itemize}
      \item $\obknowl{\kb}{\It} \models \neg a$, or
      \item for each clause $a \leftarrow a_1,...,a_n,\lpnot b_1,\ldots,\lpnot b_r$ in $\prog$
            there is some $i$ $(1 \leq i \leq n)$ such that either $a_i$ is false in $\cI$ or $a_i \in \Fa$
            (and since $\Fa\subseteq\Fa'$, $a_i \in \Fa'$),
            or some $j$ ($1 \le j \le r$) such that $b_{j}$ is true in $\cI$.
    \end{itemize}
    Also,
    if $\obknowl{\kb}{\ka{\kb}\setminus(I_F\cup\Fa)}\not\models a$,
    then
    $\obknowl{\kb}{\ka{\kb}\setminus(\If\cup\Fa')}\not\models a$
    by the monotonicity of first order logic.
    So $a\in\opfalse{\kb}{\cI}(\Fa')$.
  \end{proof}
\end{proposition}

\begin{proposition}\label{thm:optrue-opfalse-monotonic-in-i}
  Given an \hkbfs $\kb$, $\optrue\kb\cI$ and $\opfalse\kb\cI$ are monotonic in $\cI$,
  i.e., if $\cI$ and $\cI'$ are three-valued interpretations for $\kb$ such that $\cI \le \cI'$,
  then
  \begin{enumerate}
    \item for each $\Tr \subseteq \ka\kb$,
          $\optrue\kb\cI(\Tr) \subseteq \optrue\kb{\cI'}(\Tr)$
    \item for each $\Fa \subseteq \ka\kb$,
          and $\opfalse\kb\cI(\Fa) \subseteq \opfalse\kb{\cI'}(\Fa)$.
  \end{enumerate}
  \begin{proof}
    \begin{enumerate}
      \item
            If $a\in\optrue{\kb}{\cI}(\Tr)$,
            then
            \begin{itemize}
              \item either there is a clause $a \leftarrow a_1,...,a_n,\lpnot b_1,\ldots,\lpnot b_r$ in $\prog$'s grounding
                    such that for each $i$ ($1 \leq i \leq n$)
                    $a_i$ is true in $\cI$
                    (and then it is true in $\cI'$)
                    or $a_i \in \Tr$
                    and for each $j$ ($1 \le j \le r$)
                    $b_{j}$ is false in $\cI$
                    (and then it  is also false in $\cI'$);
                    which would ensure $a \in \optrue\kb{\cI'}(\Tr)$
              \item or $\obknowl{\kb}{\It \cup \Tr}\models a$, i.e., ${\transform{\ont}} \cup  \It \cup \Tr \models a$.
                    Since $\It \subseteq \It'$,
                    then also ${\transform{\ont}} \cup \It' \cup \Tr' \models a$ by the monotonicity of first order logic;
                    which, also, would ensure $a \in \optrue\kb{\cI'}(\Tr)$
            \end{itemize}
            So $a\in\optrue{\kb}{\cI'}(\Tr)$.
      \item If $a \in \opfalse\kb{\cI}(\Fa)$, then
            \begin{itemize}
              \item $\obknowl{\kb}{\ka\kb \setminus (\If\cup\Fa)} \not\models a$,
                    so also $\obknowl{\kb}{\ka\kb \setminus (\If'\cup\Fa)u} \not\models a$
                    because $\If' \supseteq \If$ and by the monotonicity of first order logic;
              \item and
                    \begin{itemize}
                      \item either $\obknowl{\kb}{\It} \models a$, so also  $\obknowl{\kb}{\It'} \models a$ by the monotonicity of first order logic;
                      \item or for all clauses
                            $a \leftarrow a_1,...,a_n,\lpnot b_1,\ldots,\lpnot b_r$ in $\prog$'s grounding
                            either there exists an $a_i \in \If \cup \Fa$, so $a_i \in \If' \cup \Fa$,
                            or a $b_j \in \It$, so $b_j \in \It'$.
                    \end{itemize}
            \end{itemize}
            In conclusion, $a \in \opfalse\kb{\cI'}(\Fa)$.
    \end{enumerate}
  \end{proof}
\end{proposition}

Given an \hkbfs $\kb$ and a 3-valued interpretation $\cI$,
since $\optrue{\kb}{\cI}$ and $\opfalse{\kb}{\cI}$ are monotonic in their argument,
they both have least and greatest fixpoints.

So it is possible to define the following iterative operator on a 3-valued interpretation $\cI$.

\begin{definition}[Iterated Fixed Point]
  \label{def:hkbfs-ifp}
  For an \hkbfs $\kb$,
  we define $\ifp\kb:\ThreeVInt\kb \rightarrow \ThreeVInt\kb$  as

  $\ifp\kb(\cI)= \pair{\lfp{\optrue\kb\cI}}{\gfp{\opfalse\kb\cI}}$.
\end{definition}

\begin{proposition}\label{thm:ifp-monotonic}
  For each \hkbfs $\kb$,  $\ifp{\kb}$ is monotonic w.r.t. the order relation among 3-valued interpretations defined in Definition \ref{def:hkbfs-3-valued-interpretation}.

  \begin{proof}

    Let $\cI$ and $\cI'$ be two three-valued interpretations of $\kb$ such that $\cI \le \cI'$.
    By propositions \ref{thm:opTrue-opFalse-monotonic} and \ref{thm:optrue-opfalse-monotonic-in-i},
    \begin{enumerate}
      \item $\upappl{\optrue\kb\cI}{n} \subseteq \upappl{\optrue\kb{\cI'}}{n}$ for all $n$
      \item $\downappl{\opfalse\kb\cI}{n} \subseteq \downappl{\opfalse\kb{\cI'}}{n}$ for all $n$
    \end{enumerate}
    Thus,
    \begin{enumerate}
      \item $\lfp{\optrue\kb\cI} \subseteq \lfp{\optrue\kb{\cI'}}$
      \item $\gfp{\opfalse\kb\cI} \subseteq \gfp{\opfalse\kb{\cI'}}$
    \end{enumerate}
    i.e., $\ifp\kb(\cI) \le \ifp\kb(\cI')$.
  \end{proof}

\end{proposition}

By virtue of being monotonic,
$\ifp\kb$ admits a least fixpoint for
each \hkbfs $\kb$,
which we define as the semantics of the \hkbfs.

\begin{definition}[Iterated fixpoint semantics]\label{def:iterated-fixpoint-semantics}
  Given an \hkbfs $\kb$, its iterated fixpoint semantics is $\lfp{\ifp\kb}$.
\end{definition}

\begin{example}[Spillover cont.]
  \label{ex:mutanp-cont}
  Consider the $\kb = \kbpair$ of Example~\ref{ex:mutanp}.
  Figure~\ref{fig:pn-ex-muta} shows the computation of the iterated fixpoint semantics for the HKB $\kb$.
  Given the presence of the function symbol $\mi{s}(\cdot)$, the model is infinite because there are countably many substitutions for $\mi{spillover\_count}$.
  \begin{figure}[htbp]
    \begin{equation}
      \begin{array}{rlrl}
        {\It}_0\ = & \emptyset                                        & {\If}_0\ = & \emptyset                                           \\
        {\It}_1\ = & \{\mi{virus}(\mi{t}),                            & {\If}_1\ = & \ka\kb \setminus {\It}_1 \setminus \{\mi{safe}(t)\} \\
        % & \mi{mutation}(\mi{t},Y),          		    &                 &          			\\
                   & \mi{mutated}(\mi{t}),                            &            &                                                     \\
                   & \mi{spillover\_count}(\mi{t},0),                 &            &                                                     \\
                   & \mi{spillover\_count}(\mi{t},\mi{s}(0)),         &            &                                                     \\
                   & \mi{spillover\_count}(\mi{t},\mi{s}(\mi{s}(0))), &            &                                                     \\
                   & \cdots\}                                         &            &                                                     \\
        {\It}_2\ = & {\It}_1                                          & {\If}_2\ = & \ka\kb \setminus {\It}_1                            \\
        {\It}_3\ = & {\It}_2                                          & {\If}_3\ = & {\If}_2	\notag
      \end{array}
    \end{equation}
    \caption{Iterations of the $\ifp\kb$ operator
      for Example~\ref{ex:mutanp}.}\label{fig:pn-ex-muta}
    % \begin{equation}
    %			\begin{array}{rlrl}
    %     \posatoms{0}\ = & \emptyset                         		    & \negatoms{0}\ = & \ka{\kb}										\\
    %     \posatoms{1}\ = & \{\mi{virus}(\mi{t}),			    		    & \negatoms{1}\ = & \{\mi{virus}(\mi{t}),              				\\
    %     & \mi{mutation}(\mi{t},Y),          		    &                 & \mi{mutation}(\mi{t},Y),            			\\
    %     & \mi{mutated}(\mi{t}),             		    &                 & \mi{mutated}(\mi{t}),               			\\
    %     & \mi{spillover\_count}(\mi{t},0),  		    &                 & \mi{spillover\_count}(\mi{t},0),   				\\
    %     & \mi{spillover\_count}(\mi{t},s(0)),   		&                 & \mi{spillover\_count}(\mi{t},s(0)),				\\
    %     & \cdots                       					&                 & \cdots  										\\
    %     \posatoms{2}\ = & \posatoms{1}\ = \negatoms{1}\ =\ \posatomsf   & \negatoms{2}\ = & \negatoms{1}\ =\ \posatoms{1}\ =\ \negatomsf	\notag
    %   \end{array}
    % \end{equation}
    % \caption{Building of $\posatomsf$ and $\negatomsf$
    % for Example~\ref{ex:mutanp}, step by
    % step.}\label{fig:pn-ex-muta}
  \end{figure}

  Each $\cI_m$, for $m = 1, 2, 3$ is determined by the fixpoints of $\optrue{\kb}{\cI_{m-1}}$ and $\opfalse{\kb}{\cI_{m-1}}$ as follows.

  \begin{itemize}
    \item $\upappl{\optrue\kb{\cI_0}}{0} = \emptyset$,
    \item $\upappl{\optrue\kb{\cI_0}}{1} = \upappl{\optrue\kb{\cI_0}}{0} \cup \{\mi{virus}(t), \mi{mutated}(t) \}$,
    \item $\upappl{\optrue\kb{\cI_0}}{2} = \upappl{\optrue\kb{\cI_0}}{1} \cup \{\mi{spillover\_count}(t,0)\}$,
    \item $\upappl{\optrue\kb{\cI_0}}{3} = \upappl{\optrue\kb{\cI_0}}{2} \cup \{\mi{spillover\_count}(t,s(0))\}$,
    \item $\upappl{\optrue\kb{\cI_0}}{4} = \upappl{\optrue\kb{\cI_0}}{3} \cup  \{\mi{spillover\_count}(t,s(s(0)))\}$,
  \end{itemize}
  and so on to the least fixpoint ${\It}_1$.

  \begin{itemize}
    \item $\downappl{\opfalse{\kb}{\cI_0}}{0} = \ka\kb$
    \item $\downappl{\opfalse{\kb}{\cI_0}}{1} = \downappl{\opfalse{\kb}{\cI_0}}{0} \setminus  \{\mi{virus}(t), \mi{mutated}(t), \mi{safe}(t) \}$
    \item $\downappl{\opfalse{\kb}{\cI_0}}{2} = \downappl{\opfalse{\kb}{\cI_0}}{1} \setminus \{\mi{spillover\_count}(t,0)\}$
    \item $\downappl{\opfalse{\kb}{\cI_0}}{3} = \downappl{\opfalse{\kb}{\cI_0}}{2} \setminus \{\mi{spillover\_count}(t,s(0))\}$
    \item $\downappl{\opfalse{\kb}{\cI_0}}{4} = \downappl{\opfalse{\kb}{\cI_0}}{3} \setminus \{\mi{spillover\_count}(t,s(s(0))) \}$
  \end{itemize}
  and so on to the greatest fixpoint ${\If}_1$.

  \begin{itemize}
    \item $\upappl{\optrue\kb{\cI_1}}{0} = \emptyset$
    \item $\upappl{\optrue\kb{\cI_1}}{1} = {\It}_1$
  \end{itemize}

  which is the least fixpoint.

  \begin{itemize}
    \item $\downappl{\opfalse{\kb}{\cI_1}}{0} = \ka\kb$
    \item $\downappl{\opfalse{\kb}{\cI_1}}{1} = \downappl{\opfalse{\kb}{\cI_1}}{0} \setminus  \{\mi{virus}(t), \mi{mutated}(t) \}$. In this case, $\mi{safe}(t)$ is kept because $\mi{spillover\_count}(s(s(0)))$ is false in $\cI_1$.
    \item $\downappl{\opfalse{\kb}{\cI_1}}{2} = \downappl{\opfalse{\kb}{\cI_1}}{1} \setminus \{\mi{spillover\_count}(t,0)\}$
    \item $\downappl{\opfalse{\kb}{\cI_1}}{3} = \downappl{\opfalse{\kb}{\cI_1}}{2} \setminus \{\mi{spillover\_count}(t,s(0))\}$
    \item $\downappl{\opfalse{\kb}{\cI_1}}{4} = \downappl{\opfalse{\kb}{\cI_1}}{3} \setminus \{\mi{spillover\_count}(t,s(s(0))) \}$
  \end{itemize}
  to the greatest fixpoint ${\If}_2 = \ka\kb \setminus {\It}_1 \cup \{\mi{safe}(\mi{t})\}$.

  For all $m$,
  $\upappl{\optrue{\kb}{\cI_2}}{m} = \upappl{\optrue{\kb}{\cI_1}}{m}$
  and $\downappl{\opfalse{\kb}{\cI_2}}{m} = \downappl{\opfalse{\kb}{\cI_1}}{m}$,
  so $\cI_2 = \cI_3$ = \lfp{\ifp\kb}.

\end{example}

\section{Properties}\label{sec:properties}

In this section,
we prove that,
for function-free {\hkbfs}s, which are also HKBs,
Knorr et al.'s alternating fixpoint partition (def. \ref{def:alternating-fixpoint-partition})
and our iterated fixpoint (def. \ref{def:iterated-fixpoint-semantics})
coincide, modulo a set complement operation.

\begin{theorem}\label{thm:hkbfs-extend-hkb}

  Given a function-free \hkbfs $\kb=\kbpair$, let $\lfp{\ifp\kb} = \pair\It\If$.
  Then $\pair\It{\ka\kb \setminus \If}$ is $\kb$'s alternating fixpoint partition.

  \begin{proof}
    We prove the claim by double induction.
    Since $\prog$ is function-free, its grounding is finite  so all fixpoints occur at finite ordinals and it is not necessary to consider limit ordinals.

    We show by induction that
    $\upappl{\ifp\kb} n=\langle\posatoms{n}; \ka{\kb}\setminus\negatoms{n}\rangle$.

    For $n=0$ (base case), $\ifp\kb\uparrow0
      = \pair{\emptyset}{\emptyset}$,
    while $\posatoms{0}=\emptyset$ and $\negatoms{0}= \ka{\kb}$,
    thus $\pair{\posatoms{0}}
      {\ka{\kb}\setminus\negatoms{0}} = \pair\emptyset\emptyset = \upappl{ \ifp\kb } 0$.

    For the inductive case, assume $\upappl{\ifp\kb}n = \pair{\posatoms{n}}{\ka{\kb}\setminus\negatoms{n}} = \cI=\pair\It\If$.

    We now prove that
    (1) $\lfp{\optrue\kb{\cI}}=
      \lfp{\TKop{\mknftrans{\kb}{\ka{\kb}\setminus \If}}}$
    and that
    (2) $\ka{\kb}\setminus (\gfp{\opfalse\kb{\cI}})=
      \lfp{\TKop{\mknfcohtrans{\kb}{\It}}}$.

    To prove (1), we first show by induction that
    $\upappl{\TKop{\mknftrans{\kb}{\ka{\kb}\setminus \If}}}m \subseteq \optrue\kb{\cI}\uparrow m$.

    For $m=0$ $\TKop{\mknftrans{\kb}{\ka{\kb}\setminus \If}}\uparrow 0=\emptyset$
    so $\TKop{\mknftrans{\kb}{\ka{\kb}\setminus \If}}\uparrow 0\subseteq \optrue\kb{\cI}\uparrow 0$.

    For $m+1$, let $\Tr$ be $\TKop{\mknftrans{\kb}{\ka{\kb}\setminus \If}}\uparrow m$,
    and assume that $\Tr\subseteq  \upappl{\optrue{\kb}{\cI}} {m}$.

    If $a\in \TKop{\mknftrans{\kb}{\ka{\kb}\setminus \If}}(\Tr)$, suppose $a\in\RKop{\mknftrans{\kb}{\ka{\kb}\setminus \If}}(\Tr)$.
    Then there exists a rule $a \leftarrow l_1,...,l_n$  in $\mknftrans{\prog_g}{\ka{\kb}\setminus \If}$, where $\prog_g$ is the grounding of $\prog$, with each $l_i\in \Tr$.
    This means that $\prog_g$ contains a rule
    $a \lpif l_1,...,l_n, \lpnot b_1, \ldots , \lpnot b_r$
    with $b_1,\ldots, b_r$  in $ \If$.
    So $a\in \upappl{\optrue\kb{\cI}}{m+1}$ by the definition of
    $\optrue\kb{\cI}$.
    If   $a\in \DKop{\mknftrans{\kb}{\ka{\kb}\setminus \If}}(\Tr)$
    then $\obknowl{\kb}{Tr}\models a$
    so  $a\in \optrue\kb{\cI}(\Tr)$
    by the definition of $\optrue\kb{\cI}$.

    Since
    $\upappl{\TKop{\mknftrans{\kb}{\ka{\kb}\setminus \If}}} m\subseteq \upappl{\optrue\kb{\cI}} m$, for all $m$,
    $\lfp{\TKop{\mknftrans{\kb}{\ka{\kb}\setminus \If}}}\subseteq
      \lfp{\optrue\kb{\cI}}$, so to prove (1) it is sufficient to show that
    $\lfp{\optrue\kb{\cI}}\subseteq
      \lfp{\TKop{\mknftrans{\kb}{\ka{\kb}\setminus \If}}}$.

    To this end, consider the sequence $S$ of sets defined by
    $S_0 = \It$, $S_{m+1} = \TKop{\mknftrans{\kb}{\ka\kb}\setminus \If}(S_m)$.
    Note that $S_m \subseteq S_{m+1}$ for all $m$,
    which can be proved by induction.
    For $m=0$, $\It = \TKop{\kb'}(\It)$ where $\kb'$ is a subset
     of $\mknftrans{\kb}{\ka\kb\setminus \If}$,
    so if $a \in \It$,
    then $a \in \TKop{\mknftrans{\kb}{\ka\kb}\setminus \If}(\It) = S_1$.
    For the inductive case, assume $S_{m-1} \subseteq S_m$:
    by the monotonicity of $\TKop{\mknftrans{\kb}{\ka\kb}\setminus \If}$ (because of Proposition 4 of \cite{KnorrMKNF_AI}, $\mknftrans{\kb}{\ka\kb \setminus \If}$ being positive),
    $S_m = \TKop{\mknftrans{\kb}{\ka\kb}\setminus \If}(S_{m-1}) \subseteq \TKop{\mknftrans{\kb}{\ka\kb}\setminus \If}(S_m) = S_{m+1}$.

    We now prove by induction that $\upappl{\optrue{\kb}{\cI}}{m} \subseteq S_m$.
    For $m = 0$, $\emptyset \subseteq \It$.
    Assuming the inclusion holds for a generic $m$,
    then if $a \in \upappl{\optrue{\kb}{\cI}}{(m+1)}$,
    either there is a rule   $a \lpif a_1,\ldots,a_n, \lpnot b_1, \ldots , \lpnot b_r$ in $\prog_g$
    with
    $\{a_1, \ldots, a_n\} \subseteq (\upappl{\optrue{\kb}{\cI}}{m} \cup \It) \subseteq (S_m \cup S_0) \subseteq S_m$
    and $\{b_1, \ldots, b_r\} \subseteq \If$,
    so $\RKop{\mknftrans{\kb}{\ka\kb}\setminus \If}$ can be applied to derive $a$;
    or $\obknowl{\kb}{\It \cup \upappl{\optrue{\kb}{\cI}}{m}}$ entails $a$,
    but then so does $\obknowl{\kb}{S_m}$,
    by the inductive hypothesis and because $\It \subseteq S_m$,
    so $\DKop{\mknftrans{\kb}{\ka\kb}\setminus \If}$ applies.

    Also, note that $\It\subseteq \lfp{\TKop{\mknftrans{\kb}{\ka{\kb}\setminus \If}}}$
    because $\It=\posatoms{n}$,
    $\posatoms{n}$ is the least fixpoint of a $\TKop{\cK'}$ operator where $\cK' \subseteq \mknftrans{\kb}{\ka{\kb}\setminus \If}$,
    and $\TKop\kb$ is monotonic in its (positive) HKB argument.
    In fact $\TKop{\kb'}(S )\subseteq \TKop\kb(S)$ if $\kb' \subseteq \kb$
    because,
    if $a$ is the head of a program rule of $\kb'$ whose body is true in $S$,
    that rule is also in $\kb$,
    and if $\obknowl{\kb'}{S}\models a$,
    then $\obknowl{\kb}{S}\models a$
    by the monotonicity of first order logic.

    Moreover, $S_m \subseteq \lfp{\TKop{\mknftrans{\kb}{\ka{\kb}\setminus \If}}}$ for all $m$.
    By induction: $S_0=\It \subseteq \lfp{\TKop{\mknftrans{\kb}{\ka{\kb}\setminus \If}}}$.
    Suppose $S_m\subseteq \lfp{\TKop{\mknftrans{\kb}{\ka{\kb}\setminus \If}}}$.
    Then $a \in S_{m+1}$ is the head of a rule of $\mknftrans{\kb}{\ka{\kb}\setminus \If}$
    whose body is true in $S_m$.
    By the inductive hypothesis,
    it is also true in $\lfp{\TKop{\mknftrans{\kb}{\ka{\kb}\setminus \If}}}$
    so $a \in \lfp{\TKop{\mknftrans{\kb}{\ka{\kb}\setminus \If}}}$.

    % each atom in $S_m$ has been derived by recursively applying $\TKop{\mknftrans{\kb}{\ka\kb}\setminus \If}$ to $\It$ at most $m$ times,
    % so it can be derived by applying the monotonic $\TKop{\mknftrans{\kb}{\ka\kb}\setminus \If}$ to its fixed point, a superset of $\It$, at most $m$ times.

    Thus, $\lfp{\optrue\kb{\cI}}\subseteq \lfp{\TKop{\mknftrans{\kb}{\ka{\kb}\setminus \If}}}$,
    which concludes the proof of (1).

    We prove (2) by proving that, for all $m$,
    $\TKop{\mknfcohtrans{\kb}{\It}}\uparrow m = \ka{\kb}\setminus(\opfalse\kb{\cI}\downarrow~m)$.

    For the base case of $m=0$,
    $\TKop{\mknfcohtrans{\kb}{\It}}\uparrow 0=\emptyset$ and $\opfalse\kb{\cI}\downarrow 0 = \ka{\kb}$,
    so $\TKop{\mknfcohtrans{\kb}{\It}}\uparrow 0 = \ka{\kb}\setminus(\opfalse\kb{\cI}\downarrow0)$.

    For the inductive case, $m+1$,
    let $S$ be $\TKop{\mknfcohtrans{\kb}{\It}}\uparrow m$
    and let $\Fa$ be $\downappl{\opfalse\kb{\cI}}{m}$.
    Note that, for all $m$,
    $\If \subseteq \downappl{\opfalse\kb\cI}{m}$,
    because by Proposition \ref{thm:ifp-monotonic}
    $\If \subseteq \gfp{\opfalse\kb\cI}$;
    thus, $\If \cup \Fa = \Fa$.

    By the inductive hypothesis,
    $S = \ka{\kb} \setminus (\If \cup \Fa)$.
    We now show that,
    for all $a \in \ka{\kb}$,
    $a\in \TKop{\mknfcohtrans{\kb}{\It}}(S)$
    if and only if $a \not\in \opfalse\kb{\cI}(\mi{Fa})$.

    % then $a \not \in \If$,
    % because otherwise $a \in \negatoms{n+1}$ 
    % and, since $\negatoms{n+1} \subseteq \negatoms{n}$, $a \in \negatoms{n}$,
    % but by the external inductive hypotheses $\If$ and $\negatoms{n}$ are disjoint.

    % So we need to prove $a \not\in \opfalse\kb{\cI}(\mi{Fa})$.
    % As just proved, $a \not \in \If$;
    Assume $a\in \TKop{\mknfcohtrans{\kb}{\It}}(S)$:
    if $\obknowl{\kb}{S} \models a$,
    then $\obknowl{\kb}{ \ka{\kb} \setminus  (\If \cup \mathit{Fa})} \models a$,
    so $a \not\in \opfalse\kb{\cI}(\Fa)$;
    otherwise, $\obknowl{\kb}{\It} \not\models \neg a$
    and there exists a rule $a \lpif a_1, \ldots, a_m, \lpnot b_1, \ldots, \lpnot b_n$ in $\prog_g$
    such that $\{a_1,\ldots,a_m\} \subseteq S$
    and $\{b_1, \ldots, b_n\} \cap \It = \emptyset$,
    which, by De Morgan's laws and because $S = \ka{\kb} \setminus (\If \cup \mathit{Fa} )$,
    is the negation of the fact that $\obknowl{\kb}{\It} \models \neg a$ or,
    for each rule $a \lpif a_1, \ldots, a_m, \lpnot b_1, \ldots, \lpnot b_n$ in $\prog_g$,
    $\{a_1,\ldots,a_m\} \cap (\If \cup \Fa) \ne \emptyset$
    or $\{b_1, \ldots, b_n\} \cap \It \ne \emptyset$;
    so again $a \not\in \opfalse\kb{\cI}(\mi{Fa})$.

    On the other hand,
    if $a \not\in \opfalse\kb{\cI}(\Fa)$,
    then either
    $(i)$ $\obknowl{\kb}{\ka{\kb} \setminus  (\If \cup \Fa)} \models a$
    (and, since $\ka{\kb} \setminus  (\If \cup \Fa) = S$, $\obknowl{\kb}{S} \models a$, so $a\in \TKop{\mknfcohtrans{\kb}{\It}}(S)$),
    or $(ii$) $\obknowl{\kb}{\It} \not\models \neg a$ and for a rule
    $a \lpif a_1, \ldots, a_m, b_1, \ldots, b_n$ in $\prog$'s grounding
     $\{a_1,\ldots,a_m\} \cap  (\If \cup \Fa) = \emptyset$
    (i.e., $\{a_1,\ldots,a_m\} \subseteq S$)
    and $\{b_1, \ldots, b_n\} \cap \It = \emptyset$,
    so again $a\in \TKop{\mknfcohtrans{\kb}{\It}}(S)$.
  \end{proof}
\end{theorem}

% \section{Related work}
% \label{sec:related-work}

% \cite{LiuYou2019-AFP-HKB} ?

\section{Conclusions and future work}\label{sec:conclusions-future-work}

In this paper we proposed an extension of the language of MKNF-based Hybrid Knowledge Bases to support function symbols in rules.
We extended the syntax
and proposed an iterative fixpoint semantics for the extended language.
We showed that the proposed semantics coincides with the one proposed by \cite{KnorrMKNF_AI} in the case of HKBs without function symbols,
so it is an extension of it.

%We are currently working on a probabilistic extension of HKBs with function symbols
%inspired by Sato's distribution semantics \cite{DBLP:conf/iclp/Sato95},
%which will be based on an iterated fixpoint operator structured similarly to the one defined in this paper.
%The language will be equipped with  a query answering system for knowledge bases defined in the language.
The proposed iterative fixpoint semantics also opens the way  to the introduction of probabilities in HKBs.
We are currently working on a probabilistic extension of HKBs with function symbols, inspired by Sato’s distribution semantics \cite{DBLP:conf/iclp/Sato95}, which will be based on the iterated fixpoint operator defined in this paper.
The probabilistic language of probabilistic \hkbfs will be also equipped with a query answering system, in the style of what we did in TRILL~\cite{DBLP:journals/tplp/ZeseCLBR19,ZesBelRig16-AMAI-IJ} and PITA~\cite{RigSwi11-ICLP11-IJ}, comparing our system with that of Knorr and colleagues~\cite{DBLP:conf/lpnmr/LopesKL17}.

\section*{Acknowledgements}

This research was partly supported by TAILOR, a project funded by EU
Horizon 2020 research and innovation programme under GA No 952215 and by
the ``National Group of Computing Science (GNCS-INDAM)".

% \bibliographystyle{eptcs}
% \bibliography{bibliography/journals_short,bibliography/booktitles_long,bibliography/series_springer,bibliography/series_long,bibliography/publishers_long,bibliography/bibl}

\newpage
\appendix

\section{MKNF-coherent HKBs}
\label{app:MKNF-coherent-hkbs}

We recall that the HKBs such that
the alternating fixpoint partition defines a three-valued MKNF model
are called \emph{MKNF-coherent}
\cite{LiuYou_AI2017_HKB_WFS}.

From Definition~\ref{def:mknf-coh}\footnote{MKNF-coherent HKB (Def. 10 of
  \cite{LiuYou_AI2017_HKB_WFS})}
An HKB $\kb$ is \emph{MKNF-coherent} if $(I_P, I_N)$, where
$I_P = \{ I \suchthat I \models \obknowl{\kb}{\posatomsf} \}$ and
$I_N = \{ I \suchthat I \models \obknowl{\kb}{\negatomsf} \}$, is a
three-valued MKNF model of \kb.

For MKNF-coherent HKBs, the model determined by the alternating
fixpoint partition as in Definition \ref{def:mknf-coh} is the unique
well-founded model.

From Proposition \ref{thm:unique-wfm}\footnote{Proposition 2 of
  \cite{LiuYou_AI2017_HKB_WFS}}
If $\kb$ is an MKNF-coherent HKB, then it has the unique well-founded
model
$(\{ I \suchthat I \models \obknowl{\kb}{\posatomsf} \},\{ I
  \suchthat I \models \obknowl{\kb}{\negatomsf} \})$

For an MKNF-coherent HKB $\kb$ with alternating fixpoint partition $(\posatomsf,\negatomsf)$ and $a \in \ka{\kb}$, we write $\WFM(\kb) \models a$ if $a \in \posatomsf$ and $\WFM(\kb) \models \neg a$ if $a \in \ka{\kb} \setminus \negatomsf$.

\cite{LiuYou_AI2017_HKB_WFS} show a bijection between
the three-valued MKNF models of an HKB $\kb$ and certain partitions of
$\ka\kb$, called \emph{stable partitions}. In the following, we report
the definition of stable partition and two results on stable partitions.

The definition of stable partition depends on the following evaluation
scheme of rules and logic programs w.r.t. partitions of the set of known atoms of an HKB.

In the following, let
$\kb = \kbpair$ be an HKB,
and $T$ and $F$ two subsets of $\ka{\kb}$
such that $T \cap F = \emptyset$.

\begin{itemize}
  \item A rule $r$ in $\prog$ is evaluated to a new rule as follows:
        \begin{itemize}
          \item $r \lbrack \mk, T, F \rbrack$ denotes the  rule obtained by replacing each
                positive literal $a$ in $r$ with \true if $a \in T$, with \false
                if $a \in F$, and with \undefine otherwise;
          \item $r \lbrack \mnot, T, F \rbrack$ denotes the rule obtained by replacing each
                negative literal $\lpnot a$ in $r$ with \true if $a \in F$, with
                \false if $a \in T$, and with \undefine otherwise;
          \item
                $r \lbrack T,F \rbrack$ denotes $r\lbrack \mk, T, F \rbrack \lbrack \mnot,
                  T, F \rbrack$.
        \end{itemize}
  \item an evaluated rule is simplified as follows:
        \begin{itemize}
          \item if the value of the head atom in a rule is equal to or greater
                than the value of its body, the rule is replaced by $\true \lpif$;
          \item if the value of the head atom in a rule is less than the value
                of its body, then the rule is replaced by $\false \lpif$.
        \end{itemize}
  \item A logic program is evaluated as follows:
        \begin{itemize}
          \item $\prog \lbrack \mk, T, F \rbrack$,
                $\prog \lbrack \mnot, T, F \rbrack$, $\prog \lbrack T, F \rbrack$
                denote the logic programs obtained by replacing each rule $r$ in $\prog$ with
                $r \lbrack \mk, T, F \rbrack$, $r \lbrack \mnot, T, F \rbrack$,
                $r \lbrack T, F \rbrack$, respectively;
          \item $\prog \lbrack \mk, T, F \rbrack$,
                $\prog \lbrack \mnot, T, F \rbrack$, $\prog \lbrack T, F \rbrack$
                evaluate to $\true$ if they are empty or if all of their rules are of the
                form $\true \lpif$; they evaluate to $\false$ if at least one rule is
                of the form $\false \lpif$.
        \end{itemize}
\end{itemize}

\begin{definition}[Stable partition -- Def. 11 of \cite{LiuYou_AI2017_HKB_WFS}]
  Let \kb be an HKB and $P \subseteq N \subseteq \ka{\kb}$. $(P,N)$ is a \emph{stable partition} of \kb if
  \begin{enumerate}
    \item \obknowl{\kb}{N} is satisfiable;
    \item $\forall a \in \ka{\kb}$, if $\obknowl{\kb}{P} \models a$
          then $a\in P$ and if $\obknowl{\kb}{N} \models a$ then $a \in N$;
          and $\prog \lbrack P, \ka{\kb} \setminus N \rbrack = \true $
    \item for any other partition $(P', N')$ with $P' \subseteq P$ and
          $N' \subseteq N$ where at least one of the inclusions is proper,
          $\exists a \in \ka{\kb} \setminus P' \suchthat \obknowl{\kb}{P'}
            \models a$, or
          $\exists a \in \ka{\kb} \setminus N' \suchthat \obknowl{\kb}{N'}
            \models a$, or
          $\prog \lbrack \mnot, P, \ka{\kb} \setminus N \rbrack \lbrack \mk, P',
            \ka{\kb} \setminus N' \rbrack = \false $
  \end{enumerate}
\end{definition}

\begin{definition}
  [Induced partition
    (Def. 7 and Lemma 1 of \cite{LiuYou_AI2017_HKB_WFS})]
  Let $S \subseteq \ka{\kb}$.
  An MKNF interpretation pair $(M,N)$ induces the partition $(T,P)$ of $S$
  by placing each atom $a \in S$ as follows:
  \begin{itemize}
    \item $a \in T$ if and only if
          $\forall I \in M, (I, ({M},{N}), ({M},{N}))(a) = \true$
    \item $a \not\in P$ if and only if
          $\forall I \in M, (I, ({M},{N}), ({M},{N}))(a) = \false$
    \item $a \in P \setminus T$ if and only if
          $\forall I \in M, (I, ({M},{N}), ({M},{N}))(a) = \undefine$
  \end{itemize}
\end{definition}

The following result establishes the correspondence between
an HKB's three-valued models
and the stable partitions of its known atoms.

\begin{theorem}[Theorem 1 of \cite{LiuYou_AI2017_HKB_WFS}]
  \label{thm:thm1}
  Let $\kb =(\ont,\prog)$ be a hybrid MKNF KB.
  \begin{itemize}
    \item If an MKNF interpretation pair $(M,N)$ is a three-valued MKNF model of $\kb$, then the partition $(T,P)$ of $\ka{\kb}$ induced by $(M,N)$ is a stable partition of $\cK$.
    \item If a partition $(T,P)$ is a stable partition of $\kb$, then the interpretation pair $(M,N)$, where $(M,N) = (\{I\suchthat I\models\obknowl{\kb}{T}\},\{I\suchthat I\models\obknowl{\kb}{P}\})$, is a three-valued MKNF model of $\kb$.
  \end{itemize}
\end{theorem}

The following theorem shows that,
for certain HKBs,
the alternating fixpoint partition is stable,
so it defines a three-valued model
which, by Theorem \ref{thm:thm1},
is the HKB's unique well-founded model.

\begin{theorem}[Theorem 3 of \cite{LiuYou_AI2017_HKB_WFS}]
  \label{thm:thm3}
  Let $\kb =(\ont,\prog)$ be a hybrid MKNF KB.
  \begin{itemize}
    \item Assume $\transform{\ont}$ is satisfiable. Then, for any $E\subseteq\ka{\kb}$, $(E,E)$ is a stable partition of $\ka{\kb}$ iff $E=\trans{\kb}{E}=\transp{\kb}{E}$.
    \item Assume $\kb$ is MKNF-coherent. Then, for any partition $(T,P)$ of $\ka{\kb}$, $(T,P)$ is a stable partition of $\kb$ iff $T = \trans{\kb}{P}$ and $P = \transp{\kb}{T}$.
  \end{itemize}
\end{theorem}

\end{document}